\documentclass[prb,superscriptaddress,showpacs,twocolumn]{revtex4}

\usepackage{graphicx}

\usepackage{amssymb,amsmath}
\usepackage{subfigure}
\usepackage[dvips]{color}

\newcommand{\ie}{i.e.\ }

\newcommand{\etal}{{\it et al.\ }}

\def\egp{\ensuremath{e_g^\pi}}

\def\t2g{\ensuremath{t_{2g}}}
\def\a1g{\ensuremath{a_{1g}}}

\def\dvo2{\ensuremath{d_{\hbox{\tiny VO}_2}}}
\def\dtio2{\ensuremath{d_{\hbox{\tiny TiO}_2}}}
\def\dsio2{\ensuremath{d_{\hbox{\tiny SiO}_2}}}

\newcommand{\eref}[1]{Eq.~(\ref{#1})}
\newcommand{\fref}[1]{Fig.~\ref{#1}}

\newcommand{\pr}{%
        ^\prime}

\newcommand{\svek}{%
        \mathbf}
\newcommand{\vek}[1]{%
        \hbox{\textbf #1}}
        
\newcommand{\fermi}[1]{%
        \hbox{f($#1$)}}
\newcommand{\dif}{%
        \hbox{d}}           
\newcommand{\tr}{%
        \hbox{ tr}}   
        
\newcommand{\im}{%
           \imath}
\newcommand{\bra}[1]{\ensuremath{\langle #1|}}
\newcommand{\ket}[1]{\ensuremath{|#1\rangle}}

\newcommand{\op}[1]{%
        \hbox{\textbf #1}}

\begin{document}

% Title of the article
\title[Optical Properties of Correlated Materials]{Optical Properties of Correlated Materials --\\ or Why Intelligent
Windows may look Dirty\footnote{This work previously appeared as the ``Scientific Highlight of the Month'' No. 88, August 2008 of the $\Psi_k$--network.}}

% Abbreviated title for the page headers
%\titlerunning{Optical Properties of Correlated Materials}

% Authors
\author{Jan M. Tomczak}
\affiliation{Research Institute for Computational Sciences, AIST, Tsukuba, 305-8568 Japan}
\affiliation{Japan Science and Technology Agency, CREST}
\author{Silke Biermann}
\affiliation{Centre de Physique Th{\'e}orique, Ecole Polytechnique, CNRS, 91128 Palaiseau Cedex, France}

%E-mail-address of corresponding author
%\mail{e-mail
%  \textsf{jan.tomczak@polytechnique.edu}, Phone:
%  +81-(0)29-8615080 ext. 55260, Fax: +81-29-861-3171}

%\received{XXXX, revised XXXX, accepted XXXX} % do not change, will be filled in by the publisher
%\published{XXXX} % do not change, will be filled in by the publisher

%Please select four to six PACS-codes from the enclosed list (PACS.txt) or from www.aip.org/pacs)
 % For example: 71.20.Ps

%\pacs{71.27.+a}{Strongly correlated electron systems; heavy fermions}
%\pacs{78.20.-e}{Optical properties of bulk materials and thin films}
%\pacs{71.30.+h}{Metal–insulator transitions and other electronic transitions}
%\pacs{71.10.-w}{Theories and models of many-electron systems}

\begin{abstract}
% This is a macro for the typesetting of two-column text in an
% abstract. It will typeset the two arguments in \abstcol{}{} as the
% left and right column inside the abstract box. At the
% columnbreak there will be always a columnbreak (\par), so both
% columns start with a new paragraph. No automatic column height
% balancing is done.
%
% If used with a \titlefigure it will silently output both
% parameters as consecutive paragraphs.
%
% The macro is defined exclusively inside the argument of \abstract{};
% if used outside it will raise an error.
%
% Usage: \abstcol{<left column>}{<right column>}
%\abstcol{% 
Materials with strong electronic Coulomb correlations
play an increasing role in modern materials applications. 
``Thermochromic'' 
systems, 
which exhibit thermally induced changes in their optical response, provide a particularly interesting case.
The optical switching associated with the
metal-insulator transition of vanadium dioxide, for example,
has been proposed for use in numerous applications,
ranging from anti-laser shields 
to ``intelligent'' windows,
which selectively filter 
radiative heat in hot weather conditions.

Are present-day electronic structure techniques able to describe,
or -- eventually even predict -- such a kind of behavior~? How far 
are we from materials design using 
%  }{%
{\it correlated oxides}~?
These are the central questions we try to address in this
article.  
  
We review recent attempts of calculating optical properties
of correlated materials within dynamical mean field theory,
and summarize results for vanadium dioxide obtained within a 
novel scheme aiming at particularly simple and 
efficient calculations of optical transition matrix elements within 
localized basis sets.

Finally, by optimizing the geometry of ``intelligent windows'', we argue that this kind of technique can in
principle 
be used 
to provide guidance for experiments,
thus 
giving a rather optimistic answer to the above questions.

\end{abstract}
%}}

% The class file requires the standard graphicx Latex package. See the 'LaTeX
% standard graphics and color packages documentation' for more information at
% <http://tug.ctan.org/tex-archive/macros/latex/required/graphics/grfguide.pdf>.
%
% Accepted figure file formats depend on which LaTeX flavour is used.
% Classic LaTeX is always able to use Encapsulted Postscript (EPS);
% PDFLaTeX can't use this but accepts PDF, JPG, PNG, and GIF formats.
%

% See examples for implementing graphics in floating figure environments later in this file.
% If \titlefigure is given, it takes as its mandatory parameter the
% name (without extension) of some figure file.
%\titlefigure[height=3.1cm]{empty2w}
%\titlefigurecaption{%
%  This is the caption of the \emph{optional} abstract figure. If
%  there is no abstract figure here, the abstract text should be divided into both columns.}

\maketitle   % please do not remove

\section{Towards Materials Design using Correlated Electron Systems~?}

Nowadays, much effort is put into finding sustainable ways to increase the energy efficiency
of man-made processes.
Besides classical engineering, it is materials science that has the potential to vastly contribute
to reduce energy spendings.
As a prototypical example, we shall describe
a proposal
made by materials scientists geared at saving
air-conditioning costs~\cite{patent:4401690,Jorgenson_window,Babulanam:1987}.

As a function of temperature, vanadium dioxide (VO$_2$) undergoes a metal-insulator transition~\cite{PhysRevLett.3.34}.
The changes through this transition are such that,
coating a window with a thin layer of VO$_2$ 
may -- under certain conditions (see below) -- create a fenestration that is ``intelligent'' in the following sense~:
At low temperatures, when VO$_2$ is in its insulating regime, the transmission properties change only remotely with respect to the bare glass window, whereas infrared radiation is filtered when the oxide layer switches to its metallic phase in hot weather conditions.
Then, heat radiation is substantially prevented from entering the building and there is less need for energy-intensive cooling by air-conditioning. 

In geographical regions, where seasonal temperature changes are important (and this 
encompasses in particular most of the industrialized world), these ``thermochromic'' windows are clearly superior to ``static'' window coatings~\cite{1402-4896-32-4-026}, 
 that filter certain wavelengths irrespective of external conditions and which thus would increase the need for heating
 buildings during winter. 

Even if in vanadium dioxide the changes of its properties
with temperature are particularly abrupt, VO$_2$
is by no means the only material where a tiny
change in external parameters can radically modify
the physical properties. In fact, such behavior can
nearly be considered a hallmark of materials with
strong electronic Coulomb interactions%
%\footnote{Vanadium dioxide could in some sense 
%be considered an exception due 
% to the fact that it
%may be less correlated than many of the other examples~\cite{me_vo2}.}
.
The concerted behavior of electrons in 
correlated materials
causes indeed quite in general an extreme sensitivity
to external stimuli, such as
temperature, pressure or external fields. 
Heating insulating SmNiO$_3$ beyond 400~K or 
applying a pressure of just a few kbar to the Mott insulator
(V$_{1-x}$Cr$_x$)$_2$O$_3$ (x=0.01)~\cite{imada}, for example,
makes the materials undergo transitions to metallic states.
This tuneability of even fundamental properties 
is both, a harbinger for diverse technological
applications and a challenge for a theoretical description.

An increasing role  
is nowadays played by
artificial
structures, ranging from multilayers that display the giant
magnetoresistance effect (widely used in storage devices)~\cite{PhysRevLett.61.2472,PhysRevB.39.4828}
to functional surfaces, where appropriate coatings e.g.\ provide a self-cleaning
mechanism~\cite{selfcleaning}.
The huge freedom in the design,  
which concerns
not only  
parameters such as
the chemical composition, the doping and the growth conditions, but
also  
the geometry of the device (e.g.\ the layer thicknesses),
however makes the search for devices with specific electronic properties
a tedious task.
This leads us to the central question of the present article~:
Can modern first principles calculations help in the quest for 
promising materials and setups for particular
devices~?
The aim can of course not be to replace experiments, but rather to 
provide some guidance in order to
minimize expensive experimental surveys and prototypings.

The Achilles heel 
of electronic structure theory is the description
of electronic many-body interactions. Indeed, a great majority of materials
used in modern applications fall in the class of so-called 
``strongly correlated materials'' where electron-electron interactions
profoundly modify (if not invalidate) a pure band picture~\cite{imada}.
State-of-the-art first principles methods, such as density functional 
theory (DFT)~\cite{RevModPhys.71.1253}, are then no longer sufficient to predict
the physical properties of these materials.
Despite these difficulties, in this article
we give a quite optimistic view on the
above question.

In fact, 
important steps to bridge the gap between band 
structure methods and many-body physics
have been made in recent years.
For materials with moderately strong correlations, Hedin's
GW approximation~\cite{hedin} -- which has seen increasingly
sophisticated implementations in the electronic structure
context~\cite{ferdi_gw,schilfgaarde:226402,bruneval:045102,RevModPhys.74.601} 
-- has established itself as a method of choice.
For strongly correlated materials,
progress was for instance brought about by 
combining 
density functional
theory within the local density approximation~\cite{RevModPhys.61.689}
with dynamical mean field theory (DMFT)~\cite{bible}.
The resulting approach, 
dubbed LDA+DMFT~\cite{0953-8984-9-35-010,PhysRevB.57.6884}
(for reviews see~\cite{vollkot,held_psik,biermann_ldadmft}),
joins
 the accurate
description of strong local Coulomb correlations in a many-body
framework with the material-specific information provided by
state-of-the-art band theory.
It has -- over the last years --
helped to elucidate physical mechanisms at work in 
systems such as transition metals~\cite{biermann_akw1,biermann_akw2,licht_katsnelson_kotliar,braun:227601}, 
their oxides~\cite{nekrasov:155112,PhysRevLett.86.5345,PhysRevLett.90.096401,pavarini:176403,PhysRevLett.91.156402,poter_v2o3,biermann:026404,tomczak_vo2_proc,me_vo2,me_phd}
or sulphides~\cite{PhysRevLett.94.166402,lechermann:085101}, as
well as f-electron compounds~\cite{PhysRevLett.87.276403,PhysRevLett.87.276404,amadon:066402,deltaPu}.
The number of applications is nowadays too large to give a complete
list in this work, but recent reviews provide an extensive
picture~\cite{vollkot,held_psik,biermann_ldadmft,RevModPhys.78.865}.

While tremendous progress has been achieved, there remains 
a chasm between what state-of-the-art electronic structure methods can calculate and what experimentalists are actually measuring. On the one hand, the chemical complexity of many systems does simply not (yet) allow for being tackled by costly many-body techniques. On the other hand, the variety of experimental observables that are being computed for the sake of comparison is rather unsatisfactory.
It is therefore an important task  to make more experimentally
measurable quantities accessible from theoretical calculations.

Within 
dynamical mean field theory, 
emphasis is commonly put on {\it spectral} properties, and
 the evaluation of 
observables other than spectral functions is a rather new advancement in the realistic context.
Yet, 
it is rather the {\it response} behavior of correlated materials that is promising for applications.

In this article, we review some recent attempts of calculating
optical properties of correlated materials, and explore  the
implications for technological applications on the specific example 
of VO$_2$-based intelligent windows.
For further reading and some of the original work see~\cite{optic_epl,optic_prb,tomczak_v2o3_proc}

%%%%%%%%%%%%%%%%%%%%%%%%%%%%%%%%%%%%%%%%%%%%%%%%%%%%%%%%%%%%%%%%%%%%%%%%%%%%%%%%%%%%%%%%%%%%%%%%%
\section{Optical Spectroscopy}

\subsection{General formulation and some physical implications}

Numerous experimental
techniques have been devised for and applied to the study of correlated materials of ever growing complexity. Optical spectroscopy, which is the subject of this work, is, in a way, the most natural among them~: Optical detectors are sampling the response to incident light, as do our eyes,
albeit accessing frequencies, and thus phenomena, that are beyond our vision.
The technique is particularly suited to track the evolution of a
system under changes of external parameters like temperature or pressure.
 This is owing to a generally high precision, and the fact that, contrary to e.g.\ photoemission spectroscopy or x-ray experiments, results are obtained in absolute values. Especially, the existence of sum-rules (see e.g.~\cite{millis_review,dressel}) allows for a quantitative 
assessment
of transfers of spectral weight upon the progression of the system properties.
Moreover, while in photoemission the electron escape depth and thus surface effects are often an issue, the larger skin penetration depth assures that optical spectroscopy is a true bulk probe. One might add that the transition matrix elements are also better understood in optics (this is an important part in this work) than in photoemission (see e.g.~\cite{RevModPhys.75.473}). 
On the other hand, response functions are two-particle quantities that are less obvious in their interpretation than a one-particle spectrum.
An important simplification is achieved when neglecting vertex corrections.
In the framework of 
linear response theory, the optical conductivity can then be expressed as
 (for reviews see~\cite{millis_review,me_phd})
\begin{eqnarray}\label{oc}
Re\,\sigma ^{\alpha\beta} (\omega)&=&\frac{2\pi e ^2\hbar}{V}\sum_{\svek{k}}\int\dif\omega\pr\;\frac{\fermi{\omega\pr}-\fermi{\omega\pr+\omega}}{\omega} \nonumber\\
&&\times
\tr\biggl\{ A_{\svek{k}}(\omega\pr+\omega) v_{\svek{k},\alpha} A_{\svek{k}}(\omega\pr)  v_{\svek{k},\beta} \biggr\} 
\end{eqnarray}
Here,
$A_{\svek{k}}(\omega)$
is the momentum-resolved many-body spectral function, 
and different optical transitions are weighted by
the Fermi velocities 
$v_{\svek{k}, \alpha}=\frac{1}{m}\bra{\svek{k}L\pr}\mathcal{P}_\alpha\ket{\svek{k}L}$, matrix elements of the momentum operator 
$\mathcal{P}$.
Both, spectral functions and velocities are matrices
in orbital space $L$, which we will specify later on.
 The Fermi functions $\fermi{\omega}$ select the range of occupied and empty energies, respectively,
$V$ is the unit-cell volume, 
$\alpha$,$\beta$ denote cartesian coordinates, and $Re\,\sigma ^{\alpha\beta}$ is the response in $\alpha-$direction for a light polarization $E$ along $\beta$.

To get an idea about the physical content of this formula, we first consider some archetypical cases. 
At zero temperature, and for a system that is well described by
its band structure as given by  density-functional theory based methods,
the (Kohn-Sham) spectrum is the defining quantity, since vertex correction are absent%
\footnote{
There are thus two effects that DFT does not account for~: (a) 
In general, it lacks the correct  spectrum. This is mended by the formalism described above, since it uses spectral functions stemming from a 
many-body calculation. 
(b) The particle-hole interaction (vertex corrections). This is not addressed in the above formalism and remains a challenge for future work.
We note that vertex corrections vanish within DMFT (infinite dimensions) in the one-band case~\cite{PhysRevLett.64.1990}. For the current multi-orbital case, this is however an approximation.
 Within GW~\cite{ferdi_gw}, vertex corrections are included on the RPA level. However, GW {\it spectra} may not be sufficient for strongly correlated systems.}%
.
Moreover, the
 spectral functions $A(\vek{k},\omega)$ of the system become Dirac
 distributions and the trace in~\eref{oc} reads\footnote{for simplicity we assume to work in the Kohn-Sham basis, \ie the spectral function $A_{\svek{k}}(\omega)$ is diagonal.}
\begin{equation}
\sum_{m,n}\delta(\omega\pr+\omega+\mu-\epsilon_\svek{k}^m) v_{\svek{k}}^{mn} \delta(\omega\pr+\mu-\epsilon_\svek{k}^n) v_{\svek{k}}^{nm}
\end{equation}
While at finite frequencies, $\omega>0$, only inter-band transitions, $m\neq n$, can give a contribution, we see that
at zero frequency, $\omega=0$, and provided a band $\epsilon_\svek{k}^n$ is crossing the Fermi level, the response is a delta function deriving from
intra-band transitions\footnote{or transitions within degenerate bands, $\epsilon_\svek{k}^n=\epsilon_\svek{k}^m$, while $m\ne n$.}.
The latter is just the result of the fact that without interactions (electron-electron correlations or a coupling to a bosonic mode) or disorder, the lattice momentum $\vek{k}$ is a constant of motion and the current thus does not decay.

In the (effective) non-interacting case 
the response of a metal to an
electric field is thus infinite. It was P. Drude in 1900 who derived
an expression that takes into account the finite lifetime of the
electron excitations, by introducing a relaxation time $\tau$ for the charge
current
$j(t)=j(0)e ^{-t/\tau}$. 
His expression for the conductivity can be
recovered from our linear response result, \eref{oc}, by
assuming free particles, $\epsilon_\svek{k}=\frac{\hbar^2 k^2}{2m}$,
and a constant self-energy\ $\Sigma=-\frac{\im}{2\tau}$.
The resulting optical conductivity is then given by
\begin{equation}
 Re\,\sigma(\omega)=\frac{ne}{m}\frac{\tau}{1+\omega ^2\tau ^2}
\end{equation}
with $n$ being the average charge carrier density. Hence, the response at zero
frequency is finite and non-zero at finite energies. 

While the physical picture about the origin of the relaxation time
had to be revised with the advent of quantum mechanics and Bloch's
theory of electronic states in periodic potentials, Drude's theory
still accounts well for the response of simple metals.
However,
it becomes 
insufficient, even in the one-band case, in the presence of substantial
correlation effects.
Indeed, the influence of electronic interactions is beyond a mere
broadening of bands. For the case of the one-band Hubbard model within
DMFT, the paramount characteristic is the appearance of Hubbard
satellites in the spectral function. Therewith, not only transitions within a broadened quasi-particle
peak are possible (\`a la Drude), but also transitions from and to 
these Hubbard
bands arise~\cite{PhysRevLett.75.105,PhysRevB.54.8452}. 
Thus, in such a  metal  two additional contributions occur,
stemming 
from transitions between the quasi-particle peak and the
individual Hubbard bands. This feature
is often referred to as the mid-infrared
peak, due to its location in energy in some compounds. At higher
energy, transitions between the two Hubbard bands appear.
In the Mott insulating phase, only the latter survive. 
Therewith the 
complexity of optical spectra is considerably enhanced~\cite{PhysRevLett.75.105,PhysRevB.54.8452}
with respect to the independent particle picture.

Within DMFT, calculations of the
optical conductivity were first performed by Pruschke and 
Jarrell \etal~\cite{PhysRevB.47.3553,PhysRevB.51.11704}
for the case of the Hubbard model.
Rozenberg~\etal~\cite{PhysRevLett.75.105,PhysRevB.54.8452} 
studied the phenomenology of the different optical
responses of the Hubbard model throughout its phase diagram,
discussing the above described phenomenology and comparing it
to experiments on V$_2$O$_3$.
With the advent of LDA+DMFT, optical conductivity calculations 
gained in realism:
Bl{\"u}mer~\etal~\cite{bluemer,blumer-2003} and later on 
Pavarini~\etal~\cite{1367-2630-7-1-188} 
and Baldassarre~\etal~\cite{Baldassarre_v2o3} 
used the
LDA+DMFT spectral functions for the calculation of titanate
and vanadate 
optical spectra.
A more general approach was developed by P{\'a}lsson~\cite{palsson} 
for the study of thermo-electricity.
Our work~\cite{me_phd,optic_epl,optic_prb,tomczak_v2o3_proc} goes along the lines of these  
approaches. We will however 
employ a formulation using the full valence Hamiltonian
therewith allowing for 
the general case including interband transitions
and we extend the intervening Fermi
velocities to multi-atomic unit cells, 
which becomes crucial in calculations for realistic compounds.

\medskip

Alternative techniques were proposed by Perlov~\etal~\cite{PhysRevB.68.245112,Ebert} 
and
 by Oudovenko~\etal~\cite{oudovenko:125112}.
The former explicitly calculated the matrix elements, albeit using
a different basis representation than in the DMFT part,
while the latter diagonalized the interacting system, which allows for
the analytical performing of some occurring integrals due to the
``non-interacting'' form of the Green's function. Owing to the
frequency-dependence of the self-energy, however, the
diagonalization has to be performed for each momentum and frequency
separately
so that the procedure may become
numerically expensive.
This technique has in particular been applied to elucidate the $\alpha-\gamma$ 
transition in Ce~\cite{haule:036401}, and, more recently,
to the heavy fermion compound CeIrIn$_5$~\cite{Shim12072007}.

The calculation of accurate absolute values of the optical response has however turned out to be a challenge
which mainly 
stems from the way the Fermi velocities are treated.
In view of predictive materials design accurate absolute values 
in the optical conductivity are a condition {\it sine qua non}.

\subsection{A scheme for the optical conductivity using
localized basis sets}

Correlation effects enter the calculation of the optical conductivity, \eref{oc}, only via the spectral functions
$A_{\svek{k}}(\omega)$, while
the Fermi velocities 

\begin{eqnarray}
v_{\svek{k}, \alpha}^{L\pr L}&=&\frac{1}{m}\bra{\svek{k}L\pr}\mathcal{P}_\alpha\ket{\svek{k}L}	
	\label{vkfull}
\end{eqnarray}
are determined by the one-particle part of the system%
\footnote{While for lattice models this is only true for local density-density interactions, in the continuum formulation of a solid this requires the interaction only to be local and of two-body kind. Then the interaction part of the Hamiltonian $\vek{H}=\vek{H}_0+\vek{H}_{int}$ commutes with the position operator $\mathcal{R}_\alpha$ and thus $\mathcal{P}_\alpha=-\im m/\hbar \left[\mathcal{R}_\alpha,\vek{H}_0 \right]$. For details see~\cite{me_phd}.}%
.

While the computation of the latter is straightforward in a plane-wave setup, the 
application of many-body techniques such as LDA+DMFT, necessitates the use
of localized basis sets (see e.g.~\cite{lechermann:125120}). 
It is thus convenient to express also the Fermi velocities in terms of this basis.
Therewith, however,  the evaluation of matrix elements of the momentum operator, \eref{vkfull},
becomes rather tedious.
 The aim of the current approach is 
to employ a controlled approximation
to the full dipole matrix elements in a Wannier like basis set that is accurate enough to yield absolute values of the conductivity
to allow for a quantitative comparison with experiment.

We choose our orbital space by specifying $L=(n,l,m,\gamma)$, with the
usual quantum numbers $(n,l,m)$, while $\gamma$ denotes the 
atoms in the unit cell~:
$\ket{\svek{k}L}$ then is the 
Fourier transform of the Wannier function
$\chi_{\svek{R}L}^{\phantom{b}}(\vek{r})$ localized at 
atom $\gamma$ in the unit cell $\vek{R}$.
Extending the well-known Peierls substitution approach for
 lattice models (see the review~\cite{millis_review}) to the realistic case of multi-atomic unit cells, we find that
\begin{eqnarray}
&&v_{\svek{k},\alpha}^{L\pr L} = \\
&&\quad{\frac{1}{\hbar} \biggl(        \partial_{k_\alpha}\op{H}^{L\pr
                   L}_{\svek{k}} -\im
                   (\rho_{L\pr}^\alpha-\rho_L^\alpha)\op{H}^{L\pr
                   L}_{\svek{k}}        
 \biggr)} + {\mathcal{F}_{\op{H}}\left[\{\chi_{\svek{R}L}^{\phantom{b}} \}\right]} \nonumber
\end{eqnarray}
Here, $\rho_L$ denotes the position of an individual atom within the unit-cell.
The term in brackets, 
which is used in the actual calculations, 
is in the following referred to as the ``generalized Peierls'' term~:
While the derivative term is the common Fermi velocity, the term proportional to the Hamiltonian originates from the generalization to realistic multi-atomic unit-cells and accounts for the fact that while the periodicity of the lattice is determined by the unit-cell, the Peierls phases couple
to the real-space positions of the individual atoms.
 The correction term that recovers the full matrix element is denoted 
$\mathcal{F}$ (for its explicit form see~\cite{optic_prb,me_phd}).
The 
latter reduces to purely atomic transitions, $(\svek{R},\gamma)=(\svek{R}\pr,\gamma\pr)$,
in the limit of strongly localized orbitals. In other words, the accuracy of the approach is controlled by the 
localization of the basis functions.
This generalized Peierls approach 
has in particular been shown to yield a good approximation 
 for systems with localized orbitals, such as the 3d or 4f orbitals
in transition metal or lanthanide/actinide compounds~\cite{me_phd}.

%%%%%%%%%%%%%%%%%%%%%%%%%%%%%
\begin{figure}[!h]	
\begin{center}
	\includegraphics[angle=0,width=0.35\textwidth]{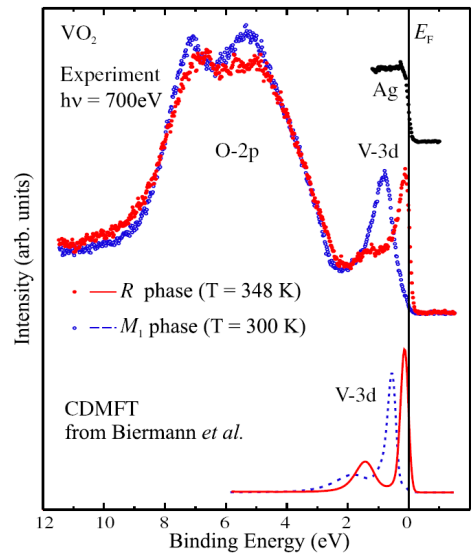}
	\caption{Comparison of valence band photoemission spectra of Koethe~\etal~\cite{koethe:116402} (top) and LDA+DMFT results from~\cite{biermann:026404} (bottom). Picture from~\cite{koethe:116402}.}
	\label{koethe}
\end{center}
\end{figure}

\section{Application~: 
 From Bulk Vanadium Dioxide to Intelligent Windows
}

\subsection{A brief reminder about the electronic structure of VO$_2$}

The electronic structure of VO$_2$ and its metal-insulator transition has been the subject of numerous theoretical
studies. We give here only a brief list of prior work, for reviews see~\cite{eyert_vo2,me_phd}.
In particular, we here do not enter at all in the decade-long discussion
on whether the metal-insulator transition in VO$_2$ should be considered
as a Peierls- or a Mott transition. Our point of view on this question 
is summarized in~\cite{me_vo2}.

\begin{figure*}[!t]
\subfigure[optical conductivity of metallic VO$_2$]{%
{\scalebox{0.48}{\includegraphics[angle=-90,width=\textwidth]{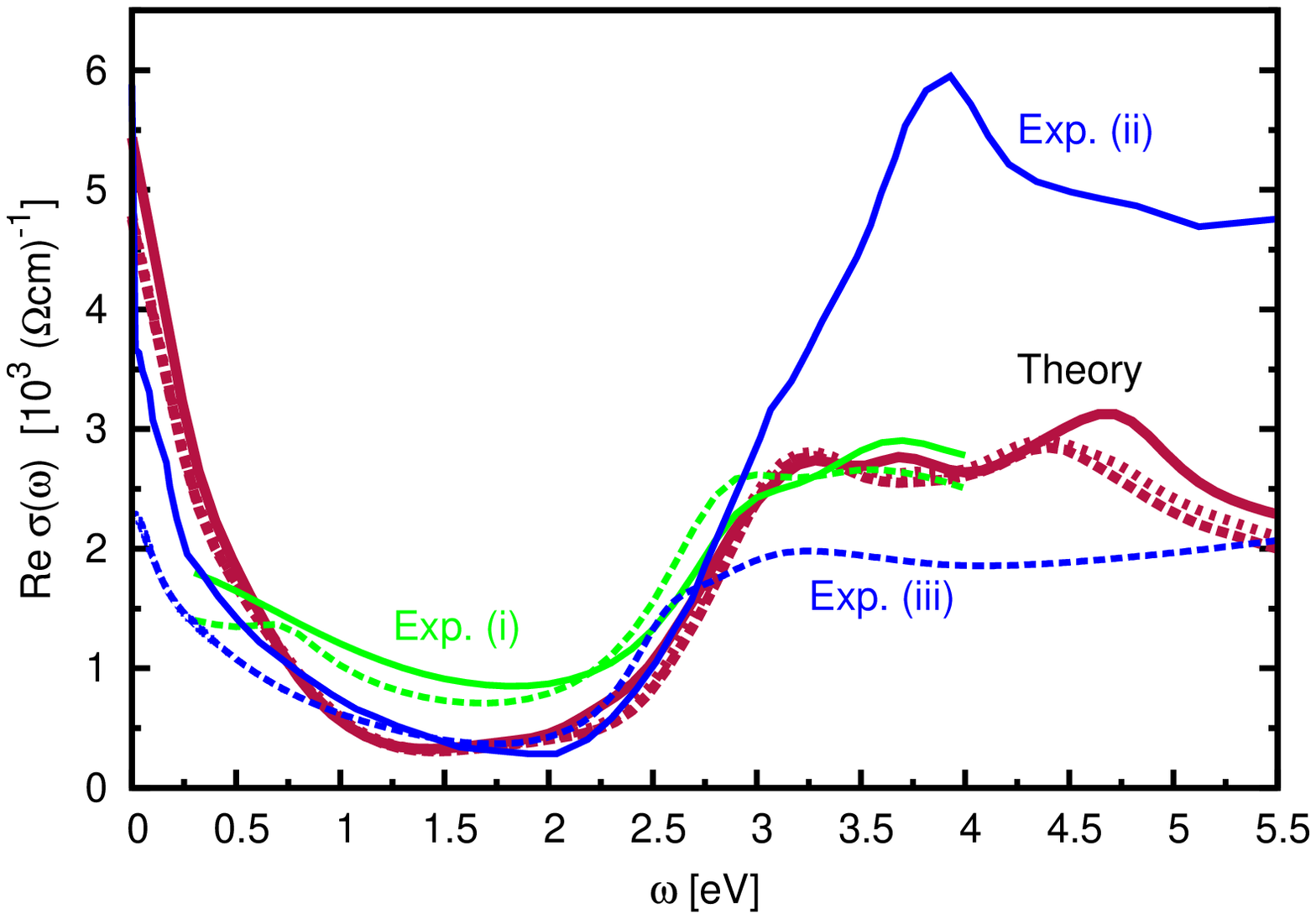}}}}\hfill
\subfigure[optical conductivity of insulating VO$_2$]{%
{\scalebox{0.48}{\includegraphics[angle=-90,width=\textwidth]{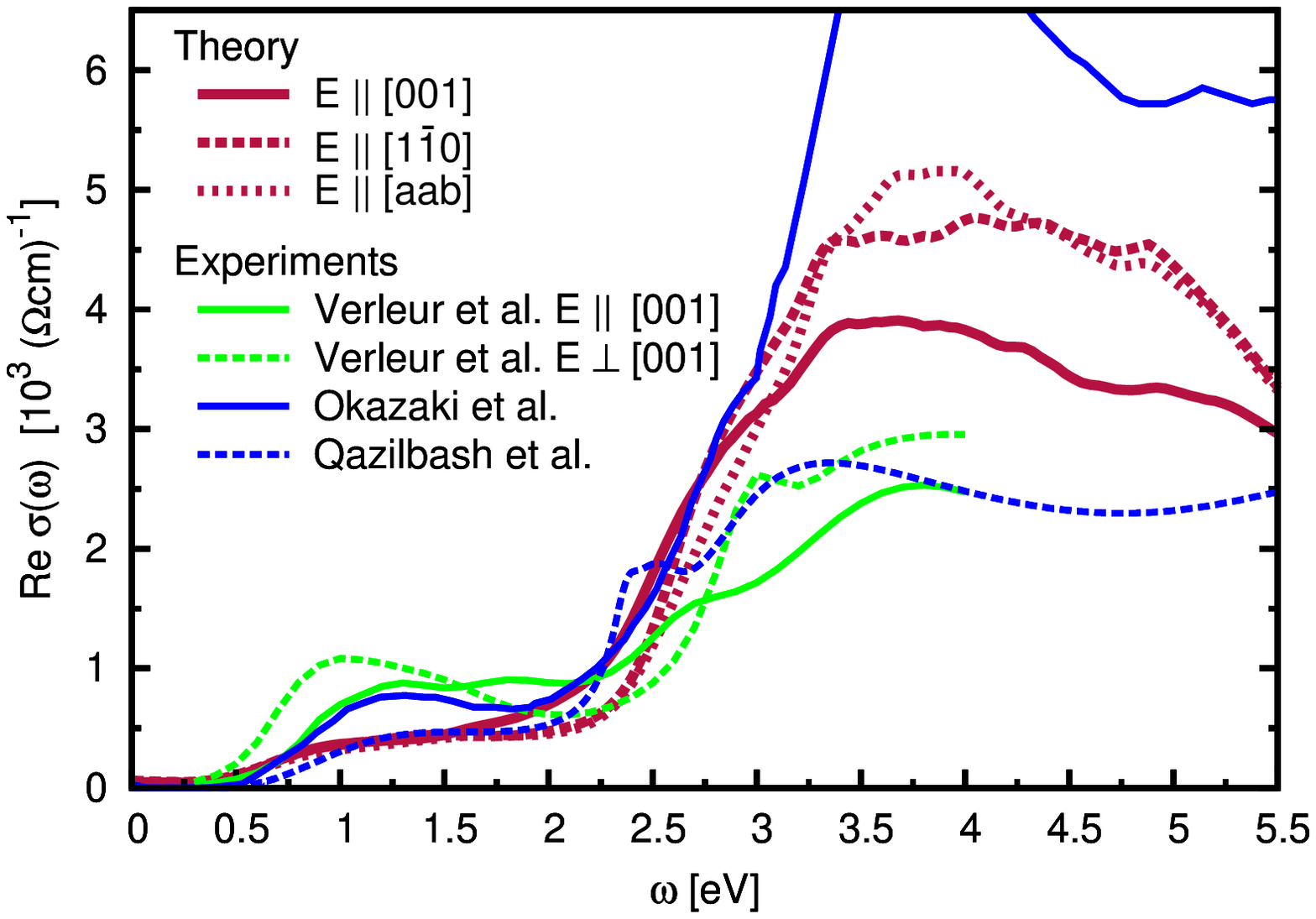}}} }\hfill
\caption{%
Optical conductivity of a) metallic, b) insulating VO$_2$ for polarizations $E$. Theory (red)
    ([aab]=[0.85 0.85 0.53]), experimental data (i)
      single crystals~\cite{PhysRev.172.788}
       (green), (ii) thin film~\cite{PhysRevB.73.165116}
       (solid blue), (iii)
      polycrystalline film~\cite{qazilbash:205118}
       (dashed blue).}
\label{fig1}
\end{figure*}

While capturing structural properties surprisingly well~\cite{PhysRevLett.72.3389},
band-structure methods 
do not reproduce experimental spectra:
in the metal, incoherent weight at higher binding energies
is absent, 
and in the insulator the gap is not opened in the corresponding
Kohn-Sham spectra.
Nevertheless, thorough LDA studies~\cite{eyert_vo2,PhysRevLett.72.3389} 
gave useful indications and paved the road to 
the application of more sophisticated many-body approaches.

LDA+DMFT results for the spectral properties of VO$_2$
agree well with experimental findings in
both, the metallic~\cite{0295-5075-69-6-984,liebsch:085109,biermann:026404} 
and the insulating phase~\cite{biermann:026404}, as can be seen from \fref{koethe} which compares photoemission
and LDA+DMFT spectra.

Since at least the insulating phase of VO$_2$ is in fact relatively
band-like~\cite{me_vo2}, an interesting alternative approach
is provided by the
GW approximation~\cite{hedin,ferdi_gw}. 
While pioneering early work~\cite{PhysRevB.60.15699} 
had to resort to a simplified scheme, recently it became possible to 
perform fully {\it ab initio} GW calculations for
 VO$_2$~\cite{me_phd,gatti:266402,Sakuma:arXiv0804.0990}. 
Full GW calculations for the optical conductivity and the spectral
function thus seem to come into reach, opening the way to
systematic comparisons with DMFT results.

Our calculation of optical properties~\cite{me_phd,optic_epl} is footing on the 
LDA+DMFT electronic structure of Ref.~\cite{biermann:026404} and 
our recent extension thereof~\cite{tomczak_vo2_proc,me_vo2}.
Since this many-body calculation used a downfolded Hamiltonian, 
we use an upfolding scheme to include optical transition from, to 
and between higher energy orbitals (For details see~\cite{tomczak_v2o3_proc}).

\subsection{The bulk conductivity}
In
\fref{fig1}(a) we display the 
LDA+DMFT 
theoretical optical conductivity of the high temperature phase of VO$_2$
 as a function of frequency and in comparison with several
experimental data~\cite{PhysRev.172.788,qazilbash:205118,PhysRevB.73.165116},
 see also~\cite{Qazilbash12142007}.
As can be inferred from the crystal structure~\cite{eyert_vo2}, 
the optical response
depends only weakly on the polarization of the incident light. 
The Drude-like metallic response (see also above) is caused by transitions between narrow vanadium 3d
orbitals near the Fermi level. As a consequence, it only influences the low
infra-red regime -- a crucial observation as we shall see in the following. The shoulder at 1.75~eV 
yet stems from intra-vanadium 3d contributions, while transitions involving 
oxygen 2p orbitals set in at 2~eV, and henceforth constitute
the major spectral weight up to the highest energies of the calculation.

In \fref{fig1}(b) we show the conductivity for
insulating VO$_2$.
As was the case before,
the theoretical results are in good agreement with the different experiments. This
time, a slight polarization dependence is seen in both, experiment and
theory,
owing to the change in 
crystal symmetry. Indeed,  vanadium
atoms pair up in the insulator to form dimers along the c-axis,
leading to the formation of bonding/anti-bonding states for 3d
orbitals\footnote{We stress that the bonding/anti-bonding splitting is not sufficient to open a gap within band-structure
approaches. Indeed correlations enhance this splitting
(after having transfered the \egp\ charge into the \a1g band), 
and we refer to this scenario as a 
``many-body Peierls insulator''~\cite{me_vo2}.%
}\cite{goodenough_vo2}. 
Optical transitions between these orbitals result in an
 amplitude of the conductivity that, in the corresponding energy range ($\omega=1.5-2.5$~eV), is higher for a light polarization parallel
 to the c-axis than for other directions (For a detailed discussion see~\cite{me_phd,optic_prb}).

\subsection{Intelligent windows}

\begin{figure*}[t]
\centering
\subfigure[setup of a VO$_2$-coated window]{%
\mbox{\centering
{\label{fig:2a}{\includegraphics[angle=-90,width=0.45\textwidth]{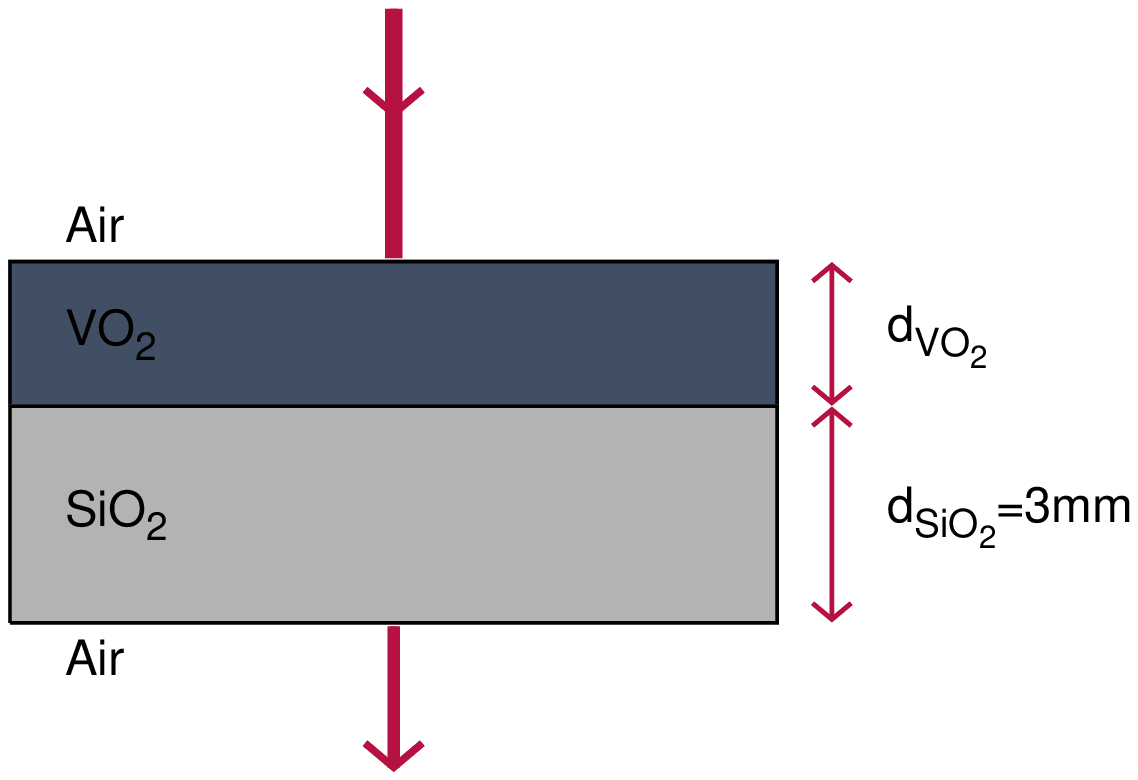}} }}}\hfill
\subfigure[reflectivity of the setup at high and low temperature]{%
{\label{fig:2b}\includegraphics[angle=-90,width=0.45\textwidth]{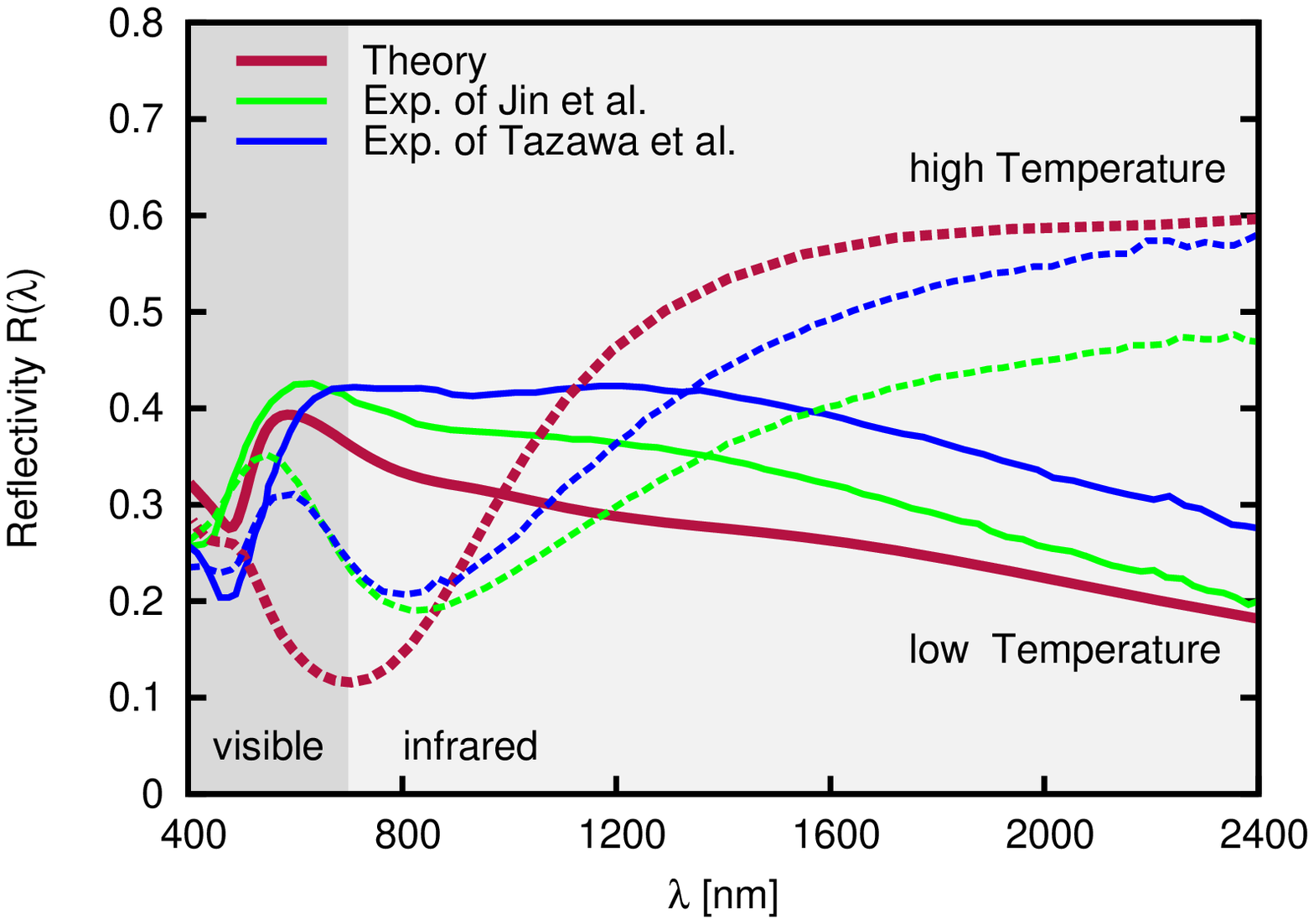}}}\hfill
\caption{%
Setup of an intelligent window.
    (a) geometry of the setup~: VO$_2$ on SiO$_2$.
    (b) Specular reflectivity
      at high (dashed) and low temperatures (solid).
      Theory (red) : 60 nm VO$_2$ on
      SiO$_2$. Experiments :
      50 nm VO$_2$ on SiO$_2$~\cite{jinvo2}
       (green), 50 nm VO$_2$ on Pyrex glass~\cite{Tazawa:98}  (blue).
       All theoretical data for $E \parallel [1\bar{1}0]$ polarization, and a 3mm SiO$_2$ substrate.
}
\label{fig2}
\end{figure*}

Having established the theoretical optical response of
{\it bulk} VO$_2$, and thus verified that our scheme can {\it quantitatively} reproduce optical properties of correlated materials, we now investigate the
possibilities of VO$_2$-based intelligent
window coatings~\cite{Jorgenson_window,Babulanam:1987,windowvo2,Chain:91}. 
The effect to be exploited here can already be seen
 in the above responses of the bulk~:
The respective conductivities of both phases (\fref{fig1}(a), (b))
exhibit a close similarity in the
range of visible light ($\omega=1.7-3.0$~eV), whereas
in the infra-red regime ($\omega<1.7$~eV) 
 a pronounced switching occurs across the metal-insulator transition.
As a result, heat radiation 
will be let through at low external
temperatures, while 
its transmission will be hindered
above the transition, which reduces the need for air conditioning in e.g.\ office buildings.
The insensitivity to temperature for visible light, in conjunction with the selectivity of the response to infrared radiation, is an
 essential feature of an intelligent window setup.
Yet, for an applicable realization, other important
 requirements have  to be met.
First of all, the switching of the window 
 has to occur at a relevant, i.e.\ ambient, temperature.
Also, the total transmittance of VO$_2$-films needs improvement in the visible range~\cite{Jorgenson_window,Babulanam:1987}, 
and the visible transmission should 
be rather independent of the wavelength
such as to provide a
colorless vision.
Experimentalists have addressed these issues and have proposed potential solutions~\cite{transcond}~:
Diverse dopings, $M_xV_{1-x}O_2$, were proven to influence the transition temperature, with Tungsten ($M=W$) being the most efficient~: A doping of only 6\%
resulted in T$_c\approx 20^\circ$C~\cite{Sobhan}.
 However, this causes a
deterioration of the infrared switching. Fluorine doping, on the other hand, improves on the switching properties, while also reducing T$_c$~\cite{tfcd,tfcd2}. 
An increase in the overall visible transmittance can also be achieved
without modifying the intrinsic properties of the material itself,
but 
by adding
antireflexion coatings with for example 
TiO$_2$~\cite{jinvo2}.

\begin{figure*}[t]%
\subfigure[setup of a multi-layer window]{%
{\hspace{-0.6cm}\label{fig:3a}\includegraphics[angle=-90,width=.35\textwidth]{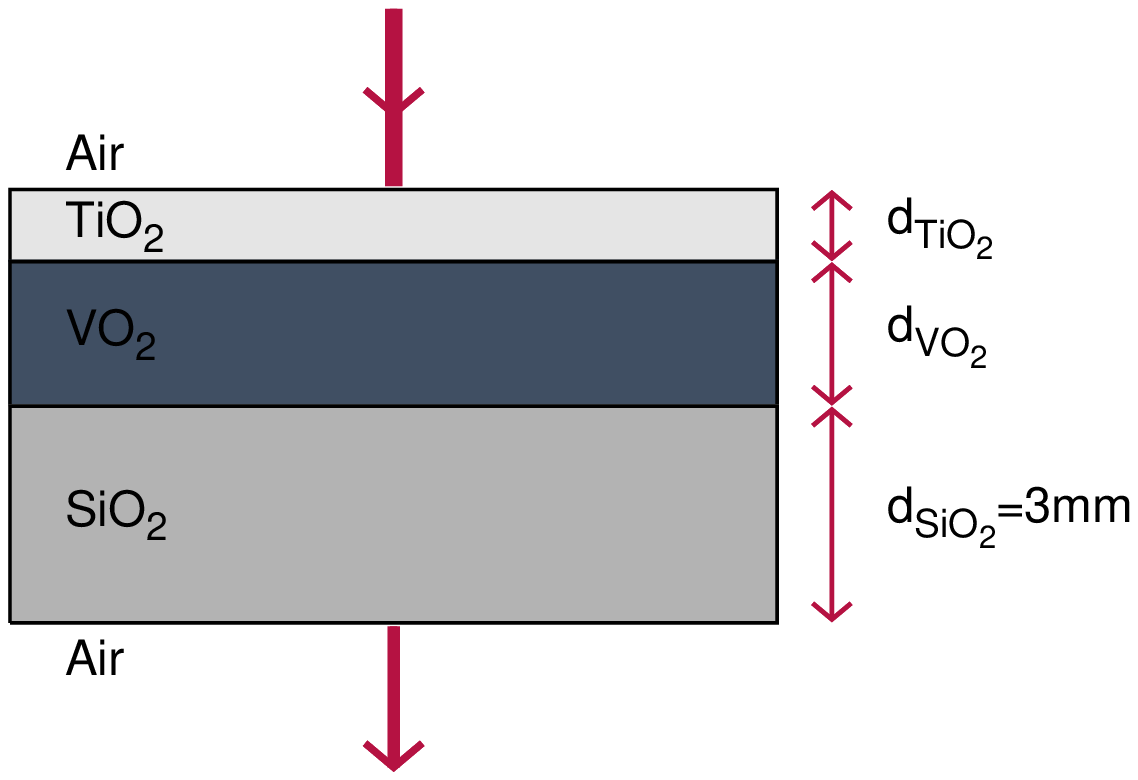}}}\hfill
\subfigure[visible transmittance for metallic VO$_2$]{%
{\hspace{-0.2cm}\includegraphics[angle=-90,width=.35\textwidth]{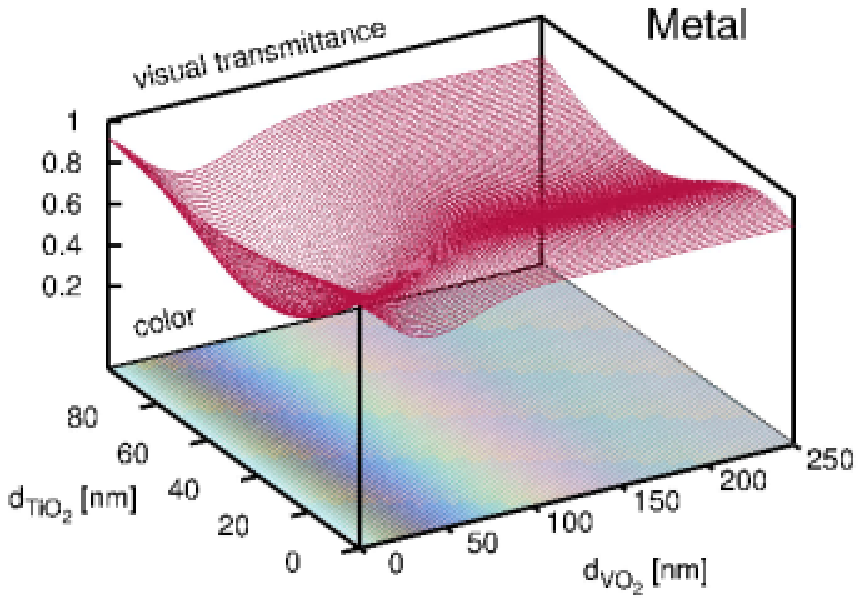}}}\hfill
\subfigure[visible transmittance for insulating VO$_2$]{%
\hspace{-0.4cm}\includegraphics[angle=-90,width=.35\textwidth]{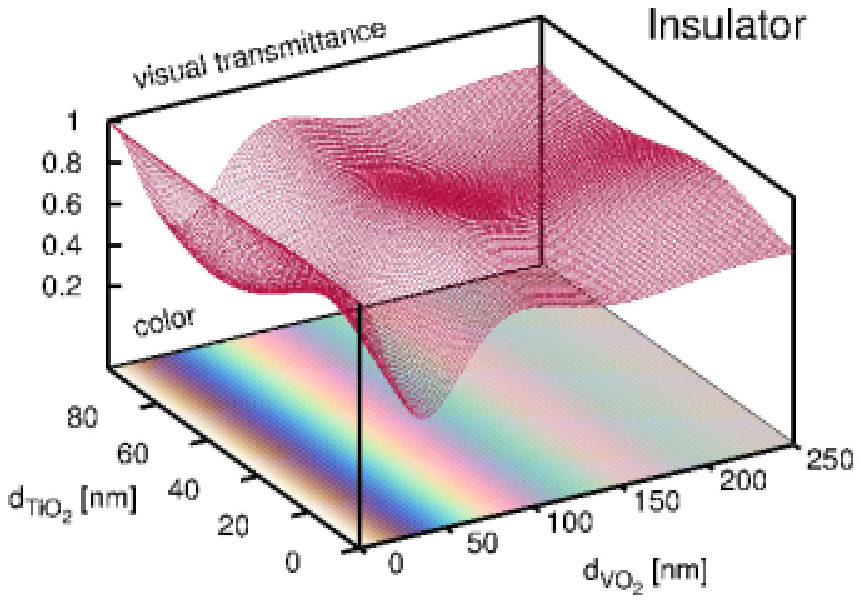}}\hfill
\caption{%
Setup of a multi-layer intelligent window.
    a) geometry of the setup~: TiO$_2$ on VO$_2$ on SiO$_2$.
       b), c) Setup with antireflexion coating : TiO$_2$ / VO$_2$ on SiO$_2$.       
   Shown is the normalized visible transmittance (see text)
   an the corresponding color of the transmitted light   
   for  b) high and c) low temperature 
   as a function of the layer thicknesses \dvo2 and \dtio2. 
All theoretical data for $E \parallel [1\bar{1}0]$ polarization, and a 3mm SiO$_2$ substrate.
}
\label{fig3}
\end{figure*}

Here, we address the optical properties of window coatings from the
theoretical perspective.
In doing so, we assume that the specular response of VO$_2$-layers
is sufficiently well-described by the optical properties of the bulk, and we use geometrical optics to deduce the properties of layered structures,
as shown e.g.\ in  \fref{fig:2a}, \fref{fig:3a}. We employ a technique equivalent to the transfer matrix method (see e.g.~\cite{tmm}), which accounts for the multiple reflexions within the layers.
The reflectivity of the setup is then given by a recursion formula. 
Defining  for the j$^{\hbox{\tiny th}}$ layer the phase
factors $\alpha_j(\omega)=\exp(\im\omega /c \hat{n}_j\delta_j)$, with $\delta_j$ and $\hat{n}_j(\omega)$ being the layer thickness and the complex refractive index, respectively, and denoting the ``bare'' reflection and transmission coefficients of an 
isolated interface between semi-infinite layers $j-1$, $j$ by r$_{j-1,j}$, t$_{j-1,j}$ one obtains a recursion formula for the ``dressed'' quantities $\tilde{r}_{0,j}$, $\tilde{t}_{0,j}$ that account for the resulting properties of the stack of the layers 0, 1, $\cdots$, j~:
\begin{eqnarray}
\tilde{r}_{0,j}&=&\tilde{r}_{0,j-1}+\tilde{t}_{0,j-1}\tilde{t}_{j-1,0}\frac{{r}_{j-1,j}\tilde{\alpha}_{j-1}^2}{1-r_{j-1,j}\tilde{r}_{j-1,0}\tilde{\alpha}_{j-1}^2}\\
\tilde{r}_{j,0}&=&{r}_{j,j-1}+{t}_{j,j-1}{t}_{j-1,j}\frac{{r}_{j-1,0}\tilde{\alpha}_{j-1}^2}{1-r_{j-1,j}\tilde{r}_{j-1,0}\tilde{\alpha}_{j-1}^2} 
\end{eqnarray}
where $\tilde{\alpha}_j=\prod_{k\le j} \alpha_k$ is an effective phase, $\tilde{t}_{j,k}=1-\tilde{r}_{j,k}$ the complex transmission, and $\tilde{r}_{j,0}$ the dressed reflectivity for light
that reaches the setup from the opposite side.

\subsubsection{A VO$_2$-coated window}

First, we consider the most simple 
possible setup, which consists of a single VO$_2$-layer (of thickness \dvo2) on a glass substrate%
\footnote{We assume the bare window to be made
of quartz glass, SiO$_2$. All auxiliary refractive indices (other than 
the one of VO$_2$) are taken from {\em Handbook of Optical Constants of Solids} by Edward~D. Palik,
Academic Press, 1985.}.
Such a window has been experimentally investigated by Tazawa \etal~\cite{Tazawa:98}
and Jin \etal~\cite{jinvo2}.
 In \fref{fig2}
  we show their measured reflectivity data 
 as a function of wavelength, in comparison with our theoretical results.

At low temperatures, i.e.\ 
insulating VO$_2$, the calculated reflectivity
is in quantitative agreement with the experimental data.
In the visible range ($\lambda\approx 400-700$~nm), the reflectivity  depends strongly on the wavelength.
Therefore, the current window
will filter certain wavelengths more than others, resulting in an illumination of a certain color
-- an obvious drawback.
 Moreover, the reflectivity in this region is rather elevated, causing
poor global transmission.
In the infrared regime ($\lambda >700$~nm) -- and beyond -- the reflectivity decreases, and heat
radiation can pass through the setup.
At high temperatures, 
the infrared reflectivity 
switches to a rather elevated value, thus filtering
heat radiation.
The changes in the visible region are less
pronounced, but still perceptible, and both, the degree
of transparency and the color  
change
 through the transition.
As a result, though presenting  qualitative features of an intelligent window, 
the current setup is not yet
 suited for applications.

%%%%%%%%%%%%%%%%%%%%%%%%%%%%%%%%%%%%%%%%%%%%%%%%%%%%%%%%%%%%%%%%%%%%%%%%%%%%%%%%%%%%%%%%%%%%%%%%%

\subsubsection{Improving the window with an anti-reflexion layer}

Having established the accuracy of our approach also for the case of layered structures,  
we now investigate a more complicated setup. 
As depicted in \fref{fig3}(a), 
an additional (rutile type) TiO$_2$-coating
is added on the VO$_2$-layer, with the objective of serving as an antireflexion filter~\cite{jinvo2}.  
With the thicknesses \dvo2 and \dtio2, the geometry of the current setup thus has
already two parameters that can be used to optimize the desired optical properties.
Since, however, each variation of them requires the production of a new individual sample 
under comparable deposition conditions, along with a careful structural
characterization in order to guarantee that differences in the optical behavior are genuine
and not related to variations of the sample quality,
the experimental expenditure is tremendous.
This led Jin \etal~\cite{jinvo2}
 to first estimate a highly transmitting setup by using tabulated
refractive indices and to produce and measure only {\it one} such sample.
Here, we shall use our theoretical results on VO$_2$ to not only optimize the geometry (\dtio2, \dvo2)
with respect to the total visible transmittance, but we shall be concerned with yet another important property
of the window that needs to be controlled~: its color.

\fref{fig3}
 displays the normalized visible specular transmittance%
\footnote{$\int_{400nm}^{700nm} d\lambda
S(\lambda)T(\lambda)/\int_{400nm}^{700nm} d\lambda S(\lambda)$, with the 
spectrum of the light
source $S(\lambda)$, and 
the transmittance $T=1-R$, $R$ being the
  specular reflexion given by Fresnel's formulae (see e.g.~\cite{dressel}). 
We thus  neglect absorption due to inhomogeneities that lead to diffuse
  reflexion. This is justified for our applications to 
 windows. Also, VO$_2$ has a glossy
  appearance and hence a preponderant specular response.}
 for our window in its high (b) and low (c) temperature state, as a function of both film thicknesses. 
 On the same graph, we moreover show the corresponding transmission color.
The evolution of the light interferences within the layers 
results in pronounced changes in both, the overall transmittance and the color.
The coating of VO$_2$ globally degrades the transparency of the bare
glass window.
An increase of the TiO$_2$-coating, on the other hand, has the potential to improve on the total
transmittance. 
This can be understood from the mechanism of commonly used
quarter-wave filters. 
The wavelength-dependence of the 
real-part of the TiO$_2$ refractive index, $n_{\hbox{\tiny TiO}_2}(\lambda)$, results in an
  optimal quarter-wave thickness,
 $\delta_{\hbox{\tiny TiO}_2}(\lambda)=\lambda/(4n_{\hbox{\tiny TiO}_2}(\lambda))$,  which varies
from blue to red light only slightly from $\delta_{\hbox{\tiny TiO}_2}(\lambda)=40$ to $60$~nm.
This and the fact that the imaginary part of the refractive index, $k_{\hbox{\tiny TiO}_2}(\lambda)$, is negligible for visible light
 also explains why the color does not change significantly with \dtio2. 
While, as for TiO$_2$, the variation of the real-part of the VO$_2$
refractive index yields a rather uniform ideal thickness $\delta_{\hbox{\tiny VO}_2}(\lambda)$, its
imaginary part changes significantly (by a factor of 4) within the range of visible light.
As a consequence, the color is very sensitive to VO$_2$-deposition. At
higher thickness \dvo2, however, this 
dependence becomes
smaller and the color lighter, as seen in \fref{fig3}(b) and (c). 
Our theoretical transmittance profiles 
suggest relatively thick windows
to yield good visual properties. 
Indeed, at low temperatures (\fref{fig3}(c)) 
the local maximum that gives the thinnest
window (\fref{fig3}(b))
 is located at $(\dtio2 , \dvo2)\approx(40~\hbox{nm}, 85~\hbox{nm})$ within our calculation.
However, this setup 
is still in the regime of important color
oscillations. Given the uncertainties  in industrial deposition techniques,
it seems rather cumbersome to consistently stabilize colorless samples.
From this point of view, a thicker VO$_2$-film would be desirable.
Indeed, while almost preserving the overall transmittance, a 
colorless window at low temperatures is realized in our calculation for $(\dtio2 , \dvo2)\approx(50~\hbox{nm},
220~\hbox{nm})$, or for $(\dtio2 , \dvo2)=(\ge 100~\hbox{nm},
220~\hbox{nm})$. 
In the high temperature state, \fref{fig3}(b),
 the transmittance is globally lower than at low temperatures.
Moreover, only the $(\dtio2 , \dvo2)=(\ge 100~\hbox{nm},
220~\hbox{nm})$ setup exhibits a simultaneously high transmittance
in {\it both} states of the window.

%%%%%%%%%%%%%%%%%%%%%%%%%%%%%%%%%%%%%%%%%%%%%%%%%%%%%%%%%%%%%%%%%%%%%%%%%%%%%%%%%%%%%%%%%%%%%%%%%
\section{Conclusions and Perspectives}

In this article,
 we have reviewed 
a scheme for optical properties of correlated materials
that is geared 
at fast and accurate calculations within localized basis sets.

As an example, we have 
addressed -- for the first time from a theoretical
perspective -- the fascinating application of the seemingly simple composition of vanadium dioxide 
in so-called intelligent windows.

Even if, from a technical point of view,
possible ways of improvement in view of
a fully general optics scheme are obvious 
-- ideally, one may want to include e.g.\ vertex corrections, 
replace the LDA+DMFT
starting point by a GW+DMFT calculation~\cite{PhysRevLett.90.086402}
or 
calculate at least the
ligand-orbital energies by many-body techniques
such as GW 
--
our simple and efficient optics scheme has
proven to reach quantitative agreement with
experiments, at least for materials where
excitonic effects are negligible.
We have shown that it is not only useful to address
questions of fundamental physics,
but that one can indeed think of exploiting 
present day techniques for applied purposes.

One may argue, of course, that a crucial ingredient
entering LDA+DMFT calculations is a reliable estimate
for the local Coulomb interactions, the Hubbard U.
In general, 
it is probably a fair statement
to say that -- despite a number of techniques that have by now
been proposed for the calculation of U
(constrained LDA~\cite{constrainedLDA}, 
constrained RPA~\cite{PhysRevB.70.195104,miyake:085122},
or GW+DMFT~\cite{PhysRevLett.90.086402,gwdmft_proc1,gwdmft_proc2})
 -- its determination still presents one of the
bottlenecks for {\it ab initio} materials
design. The recently much discussed new family
of iron pnictide superconductors
provides an interesting test case for calculations of 
U~\cite{JPSJS.77.093711,JPSJS.77SC.99,0953-8984-21-7-075602}, where we dispose of little {\it a priori}
knowledge.

For vanadium dioxide, however, this issue seems to be more a
conceptual than a practical one, since experimental
and theoretical estimates finally converge towards a common
answer.

In conclusion, we thus
 give an
optimistic answer to the central question of this article~:
electronic structure techniques can -- if not
design -- at least help guiding the search for functional 
materials and their devices, and this even for the
particularly challenging class of correlated materials.

\section{acknowledgement}
We thank the numerous colleagues, with whom we have had fruitful
discussions, collaborations and exchanges on different 
aspects related to this research, in particular~:
O.K. Andersen,
F. Aryasetiawan, L. Baldassarre, T. Gacoin, G. Garry, M. Gatti,  A. Georges, K. Haule,
H.J. Kim, 
G. Kotliar, A.I. Lichtenstein, M. Marsi, T. Miyake, 
V. Oudovenko, A.I. Poteryaev, J.P. Pouget, L. Pourovskii,
M.M. Qazilbash, L. Reining, F. Rodolakis, R. Sakuma,
 E. Wimmer, R. Windiks.

This work was supported by  
the French ANR under project
CORRELMAT, and a computing grant by IDRIS Orsay under project number
091393.
%\end{acknowledgement}

% Use the following code if you wish to generate your bibliography with BibTeX;
% replace the string "pss-demo" below with the name(s) of
% the BibTeX data base(s) you want to use.
% The resulting bibliography-output (the contents of the .bbl file)
% must be pasted back into this file before submission.
%
%\bibliographystyle{pss}
%\bibliography{../../../refs,../../../refs_mine}
%

% Replace the following example bibliography with your references

\end{document}